\documentclass[12pt]{article}
\usepackage{amssymb,amsfonts}
\usepackage{epsf,epsfig}
\begin{document}
\baselineskip 18pt

\title{Three magnons in an isotropic $S=1$ ferromagnetic chain as an exactly solvable non-integrable system}
\author{P.N. Bibikov}
\date{\it Russian State Hydrometeorological University, Saint-Petersburg, Russia}
\maketitle

\vskip5mm

\begin{abstract}
It is shown that a generalization of Bethe Ansatz based on an utilization of {\it degenerative discrete-diffractive} wave functions solves the three-magnon problem for the $S=1$ isotropic ferromagnetic infinite chain. The four-magnon problem is briefly discussed.
\end{abstract}

\maketitle

\section{Introduction}

Quantum integrable models most usually are solved by various versions of Bethe Ansatz \cite{1,2,3}. However solvability does not imply
integrability \cite{4}. In fact the later results in a non ergodic physical behavior, while the former gives the possibility to obtain
the spectrum of physical states. An interplay between these conceptions may be well illustrated for models whose particles (elementary excitations) are flat waves excited from the ground state $|\emptyset\rangle$ by some creation operators ${\bf\bar\Psi}_n^j$. Here the index $j=1,\dots,d$ parameterizes internal degrees of freedom such as polarization of triplons in spin ladders ($d=3$) \cite{5}, or electrons in the $t-J$ model ($d=2$) \cite{6}. Namely one- and two-particle states are
\begin{eqnarray}
&&|k\rangle^j=\sum_n{\rm e}^{ikn}{\bf\bar\Psi}_n^j|\emptyset\rangle,\\
&&|k_1,k_2\rangle^{j_1j_2}=\sum_{n_1<n_2}\Big[A_{12}^{j_1j_2}{\rm e}^{i(k_1n_1+k_2n_2)}-A_{21}^{j_1j_2}{\rm e}^{i(k_2n_1+k_1n_2)}\Big]
{\bf\bar\Psi}_{n_1}^{j_1}{\bf\bar\Psi}_{n_2}^{j_2}|\emptyset\rangle.
\end{eqnarray}
As vectors in ${\mathbb C}^d\otimes{\mathbb C}^d$ the two amplitudes $A_{12}^{j_1j_2}$ and $A_{21}^{j_1j_2}$ are related by the formula
\begin{equation}
A_{21}=S(k_1,k_2)A_{12},
\end{equation}
where the $d^2\times d^2$ matrix $S(k_1,k_2)$ is the two-particle scattering matrix. If it satisfies the Yang-Baxter equation \cite{7}
\begin{equation}
S_{12}(k_b,k_c)S_{23}(k_a,k_c)S_{12}(k_a,k_b)=S_{23}(k_a,k_b)S_{12}(k_a,k_c)S_{23}(k_b,k_c),
\end{equation}
where ($I$ is the finite-dimensional matrix unit)
\begin{equation}
S_{12}(k,\tilde k)=S(k,\tilde k)\otimes I,\quad S_{23}(k,\tilde k)=I\otimes S(k,\tilde k),
\end{equation}
then the wave function of a three-particle state
\begin{equation}
|k_1,k_2,k_3\rangle^{j_1j_2j_3}=\sum_{n_1<n_2<n_3}a_{n_1,n_2,n_3}^{j_1j_2j_3}{\bf\bar\Psi}_{n_1}^{j_1}{\bf\bar\Psi}_{n_2}^{j_2}
{\bf\bar\Psi}_{n_3}^{j_3}|\emptyset\rangle,
\end{equation}
may be obtained in the Bethe form (we have omitted the polarization indices, and used the Levi-Civita tensor)
\begin{equation}
a_{n_1,n_2,n_3}(k_1,k_2,k_3)=\sum_{a,b,c=1}^3\varepsilon_{abc}A_{abc}(k_1,k_2,k_3){\rm e}^{i(k_an_1+k_bn_2+k_cn_3)},
\end{equation}
which express both solvability and integrability. The former is evident, while the latter follows from the fact that according to (7) an initial set of incoming wave numbers does not change under the scattering.

Equation (4) was first treated as an integrability condition for quantum gas with delta-function interaction \cite{8} and then used as an
integrability test for other models such as the above-mentioned spin ladder and t-J model \cite{5,6}.
For a 1D system equation (4) together with an {\it absence of particle production} are necessary conditions
for factorization of the multi-particle scattering or equivalently for its reduction to a
succession of space-time separated two-particle collisions \cite{9}. For identical particles due to the energy and momentum conservation
laws, such a collision reduces to an exchange of wave numbers between scattering particles and multiplication of amplitudes on an appropriate
two-particle $S$-matrix (see equation (3)). That is why an $m$-particle wave function is a sum of $m!$ exponential terms each of them corresponds to a
permutation of wave numbers in a set $\{k_1,\dots,k_m\}$. Since a distribution of wave
numbers (momentums) does not alter under the collisions the system possesses a non ergodic behavior.

For a non-integrable quantum system the situation is drastically different. One- and two-particle states may be obtained as previously \cite{5,6}.
However a three-particle wave function should have a diffractive form \cite{3,10,11}
\begin{eqnarray}
&&a_{n_1,n_2,n_3}=\int dk_1dk_2dk_3\delta\Big(\prod_{l=1}^3{\rm e}^{ik_l}-{\rm e}^{ik}\Big)\delta\Big(\sum_{l=1}^3E(k_l)-E\Big)\nonumber\\
&&\cdot\sum_{a,b,c=1}^3\varepsilon_{abc}A_{abc}(k_1,k_2,k_3){\rm e}^{i(k_an_1+k_bn_2+k_cn_3)},
\end{eqnarray}
corresponding to changes of an incoming triple of wave numbers and hence to an ergodic behavior. Unfortunately a substitution of (8) into (6) does not reduce the spectral problem to a rather simple form as it did in the case (7).

In the present paper we suggest a discrete analog of the diffractive form (8), namely
\begin{equation}
a_{n_1,n_2,n_3}=\sum_{m=1}^{M}\sum_{a,b,c=1}^3\varepsilon_{abc}A_{abc}^{(m)}{\rm e}^{i(k_a^{(m)}n_1+k_b^{(m)}n_2+k_c^{(m)}n_3)}.
\end{equation}
Here $M>1$ and
\begin{eqnarray}
\prod_{l=1}^3{\rm e}^{ik_l^{(m)}}={\rm const},\quad \sum_{l=1}^3E(k_l^{(m)})={\rm const}.
\end{eqnarray}
As will be shown here for a degenerative case $M<\infty$  the ansatz (9) results in essential simplifications. Namely for
\begin{equation}
1<M<\infty,
\end{equation}
the system is still ergodic and non-integrable but the three-particle wave functions (9) may be derived exactly as in the usual version of Bethe Ansatz.

Magnons (spin waves with $\Delta S=1$) in a $S=1$ isotropic ferromagnetic chain (see the Hamiltonian (12)-(13) of the present paper) have no internal
degrees of freedom. Hence the two-magnon $S$-matrix is a scalar function and equation (4) is satisfied automatically. However the two-magnon scattering
results in the creation of the quadruplon resonance \cite{12} (the spin wave with $\Delta S=2$). In the ${\rm su}(3)$-symmetric point \cite{13} the quadruplons are
stable particles with the same energy as magnons. For a broken ${\rm su}(3)$-symmetry the quadruplons become instable and decay into magnon pairs, at the same time creating as resonances in two-magnon collisions. Due to these processes a three-magnon scattering has the
following channel. First of all two neighboring incoming magnons create a quadruplon resonance which then collide with the third incoming
magnon. Under this collision the resonance decays on two magnons. One of them goes to infinity while the other forms a new resonance with the third magnon. Finitely this new resonance again decays on two outgoing magnons.
As it will be shown in Sect. 4 only this process can not be accounted by the wave functions of the form (7).
However the $M=2$ wave functions of the form (9) will be derived.

\section{Hamiltonian and one-magnon spectrum}

In the present paper we study the general model of isotropic $S=1$ ferromagnet with the Hamiltonian
\begin{equation}
\hat H=\sum_{n=-\infty}^{\infty}H_{n,n+1},
\end{equation}
where \cite{12,14}
\begin{equation}
H_{n,n+1}=\hat I-({\bf S}_n\cdot{\bf S}_{n+1})+J\Big(\hat I-({\bf S}_n\cdot{\bf S}_{n+1})^2\Big).
\end{equation}
Here ${\bf S}_n$ is the standard triple of $S=1$ spin operators associated with $n$-th site. The constant terms proportional to the infinite-dimensional matrix
unit $\hat I$ are added only for the relation
\begin{equation}
\hat H|\emptyset\rangle=0,
\end{equation}
where the state
\begin{equation}
|\emptyset\rangle=\prod_{n=-\infty}^{\infty}\otimes|+1\rangle_n,
\end{equation}
(the states $|j\rangle_n$  with $j=-1,0,1$ form the standard $S=1$ triple associated with $n$-th site)
will be treated as the pseudovacuum. It may be readily proved by an analysis of the spectrum of the $9\times9$ Hamiltonian density matrix related to
$H_{n,n+1}$ that for $J<1$ equation (15) gives the ground state of the model. For $J>1$ it will be the ground state under the saturating magnetic field.

Using the representation
\begin{equation}
({\bf S}_n\cdot{\bf S}_{n+1})={\bf S}_n^z{\bf S}_{n+1}^z+\frac{1}{2}\Big({\bf S}_n^+{\bf S}_{n+1}^-+{\bf S}_n^-{\bf S}_{n+1}^+\Big),
\end{equation}
and relations
\begin{equation}
{\bf S}^{\pm}_n|\pm1\rangle_n=0,\quad{\bf S}^{\pm}_n|0\rangle_n=\sqrt{2}|\pm1\rangle_n,\quad
{\bf S}^{\pm}_n|\mp1\rangle_n=\sqrt{2}|0\rangle_n,\quad{\bf S}^z_n|j\rangle_n=j|j\rangle_n.
\end{equation}
one readily gets \cite{14}
\begin{eqnarray}
&&H_{n,n+1}|\pm1\rangle_n\otimes|\pm1\rangle_{n+1}=0\nonumber\\
&&H_{n,n+1}|\pm1\rangle_n\otimes|0\rangle_{n+1}=|\pm1\rangle_n\otimes|0\rangle_{n+1}
-|0\rangle_n\otimes|\pm1\rangle_{n+1},\nonumber\\
&&H_{n,n+1}|0\rangle_n\otimes|\pm1\rangle_{n+1}=|0\rangle_n\otimes|\pm1\rangle_{n+1}
-|\pm1\rangle_n\otimes|0\rangle_{n+1},\nonumber\\
&&H_{n,n+1}|0\rangle_n\otimes|0\rangle_{n+1}=(1-J)\Big(|0\rangle_n\otimes|0\rangle_{n+1}
-|1\rangle_n\otimes|-1\rangle_{n+1}-|-1\rangle_n\otimes|1\rangle_{n+1}\Big),\nonumber\\
&&H_{n,n+1}|\pm1\rangle_n\otimes|\mp1\rangle_{n+1}=(2-J)|\pm1\rangle_n\otimes|\mp1\rangle_{n+1}
+(J-1)|0\rangle_n\otimes|0\rangle_{n+1}\nonumber\\
&&-J|\mp1\rangle_n\otimes|\pm1\rangle_{n+1}.
\end{eqnarray}

A one-magnon state is the flat wave \cite{12,14}
\begin{equation}
|1,k\rangle=\sum_n{\rm e}^{ikn}{\bf S}_n^-|\emptyset\rangle.
\end{equation}
whose energy
\begin{equation}
E_{magn}(k)=2(1-\cos{k}),
\end{equation}
may be readily obtained from (18).

\section{The two-magnon scattering}

A two-magnon state should have the form \cite{14}
\begin{equation}
|2\rangle=\Big[\sum_{n_1<n_2}a_{n_1,n_2}{\bf S}_{n_1}^-{\bf S}_{n_2}^-
+\sum_nb_n({\bf S}_n^-)^2\Big]|\emptyset\rangle.
\end{equation}
According to (18) the corresponding ${\rm Schr\ddot odinger}$ equation on the combined wave function $\{a_{n_1,n_2},b_n\}$ has the form of the following system \cite{14}
\begin{eqnarray}
&&4a_{n_1,n_2}-a_{n_1-1,n_2}-a_{n_1+1,n_2}-a_{n_1,n_2-1}-a_{n_1,n_2+1}=Ea_{n_1,n_2},\quad n_2-n_1>1,\\
&&(3-J)a_{n,n+1}-a_{n-1,n+1}-a_{n,n+2}+(J-1)(b_n+b_{n+1})=Ea_{n,n+1},\nonumber\\
&&2(2-J)b_n-J(b_{n-1}+b_{n+1})+(J-1)(a_{n-1,n}+a_{n,n+1})=Eb_n.
\end{eqnarray}
In the ${\rm su}(3)$ invariant point $J=1$ the system (22)-(23) splits into two independent subsystems and a solution
\begin{equation}
b_n={\rm e}^{ikn},
\end{equation}
with the same energy as (20) corresponds to the above-mentioned quadrupole wave \cite{12,13} (quadruplon).
Turning to the general case $J\neq1$ and suggesting the following ansatz
\begin{equation}
a_{n_1,n_2}(k_1,k_2)=A_{12}{\rm e}^{i(k_1n_1+k_2n_2)}-A_{21}{\rm e}^{i(k_2n_1+k_1n_2)},\quad
b_n(k_1,k_2)=B{\rm e}^{i(k_1+k_2)n},
\end{equation}
we readily get from (22) the energy
\begin{equation}
E(k_1,k_2)=E_{magn}(k_1)+E_{magn}(k_2),
\end{equation}
and reduce (23) to
\begin{eqnarray}
&&\Big[1+{\rm e}^{i(k_1+k_2)}-(1+J){\rm e}^{ik_2}\Big]A_{12}-\Big[1+{\rm e}^{i(k_1+k_2)}-(1+J){\rm e}^{ik_1}\Big]
A_{21}\nonumber\\
&&+(J-1)\Big(1+{\rm e}^{i(k_1+k_2)}\Big)B=0,\nonumber\\
&&(J-1)\Big({\rm e}^{-ik_1}+{\rm e}^{ik_2}\Big)A_{12}-(J-1)\Big({\rm e}^{ik_1}+{\rm e}^{-ik_2}\Big)A_{21}\nonumber\\
&&+2\Big(\cos{k_1}+\cos{k_2}-J(\cos{(k_1+k_2)}+1)\Big)B=0.
\end{eqnarray}

System (27) has the following solutions
\begin{equation}
A_{12}=A(k_1,k_2),\quad A_{21}=A(k_2,k_1),\quad B=B(k_1,k_2),
\end{equation}
where
\begin{equation}
A(k,\tilde k)={\rm e}^{i\tilde k}+{\rm e}^{i(k+2\tilde k)}-(1+J){\rm e}^{2ik}-(1+3J){\rm e}^{i(k+\tilde k)}+
J\Big(3{\rm e}^{ik}+3{\rm e}^{i(2k+\tilde k)}-1-{\rm e}^{2i(k+\tilde k)}\Big),
\end{equation}
and
\begin{equation}
B(k,\tilde k)=(1-J)\Big({\rm e}^{i\tilde k}-{\rm e}^{ik}\Big)(1+{\rm e}^{i(k+\tilde k)}).
\end{equation}

As we see the two-magnon problem may be solved in a non-diffractive way for all values of $J$. In other words it is unsensitive to non-integrability. Really equation (26) together with the condition $k_1+k_2=k$ ($k$ is the total momentum of the state)  define the pair of wave numbers $k_1$ and $k_2$ up to a permutation of them.

\section{The three-magnon problem}

A three-magnon state should have the form
\begin{equation}
|3\rangle=\Big[\sum_{n_1<n_2<n_3}a_{n_1,n_2,n_3}{\bf S}_{n_1}^-{\bf S}_{n_2}^-{\bf S}_{n_3}^-
+\sum_{n_1<n_2}\Big(b^{(1)}_{n_1,n_2}({\bf S}_{n_1}^-)^2{\bf S}_{n_2}^-+b^{(2)}_{n_1,n_2}{\bf S}_{n_1}^-({\bf S}_{n_2}^-)^2\Big)\Big]|\emptyset\rangle
\end{equation}
The corresponding ${\rm Shr\ddot odinger}$ equation on the combined wave function $\{a_{n_1,n_2,n_3},b^{(1)}_{n_1,n_2},b^{(2)}_{n_1,n_2}\}$ splits on four groups of equations
\begin{equation}
6a_{n_1,n_2,n_3}-a_{n_1+1,n_2,n_3}-a_{n_1,n_2+1,n_3}-a_{n_1,n_2,n_3+1}-a_{n_1-1,n_2,n_3}
-a_{n_1,n_2-1,n_3}-a_{n_1,n_2,n_3-1}=Ea_{n_1,n_2,n_3},
\end{equation}
at $n_2-n_1>1$, $n_3-n_2>1$,
\begin{eqnarray}
&&(5-J)a_{m-1,m,n}-a_{m-1,m+1,n}-a_{m-1,m,n+1}-a_{m-2,m,n}-a_{m-1,m,n-1}\nonumber\\
&&+(J-1)\Big[b^{(1)}_{m-1,n}+b^{(1)}_{m,n}\Big]=Ea_{m-1,m,n},\nonumber\\
&&(5-J)a_{m,n,n+1}-a_{m,n,n+2}-a_{m+1,n,n+1}-a_{m,n-1,n+1}-a_{m-1,n,n+1}\nonumber\\
&&+(J-1)\Big[b^{(2)}_{m,n}+b^{(2)}_{m,n+1}\Big]=Ea_{m,n,n+1},\nonumber\\
&&2(3-J)b^{(1)}_{m,n}-b^{(1)}_{m,n+1}-b^{(1)}_{m,n-1}-J\Big[b^{(1)}_{m-1,n}+b^{(1)}_{m+1,n}\Big]\nonumber\\
&&+(J-1)\Big[a_{m,m+1,n}+a_{m-1,m,n}\Big]=Eb^{(1)}_{m,n},
\nonumber\\
&&2(3-J)b^{(2)}_{m,n}-b^{(2)}_{m+1,n}-b^{(2)}_{m-1,n}-J\Big[b^{(2)}_{m,n-1}+b^{(2)}_{m,n+1}\Big]\nonumber\\
&&+(J-1)\Big[a_{m,n,n+1}+a_{m,n-1,n}\Big]=Eb^{(2)}_{m,n},
\end{eqnarray}
at $n-m>1$,
\begin{equation}
2(2-J)a_{n-1,n,n+1}-a_{n-1,n,n+2}-a_{n-2,n,n+1}+(J-1)\Big[b^{(1)}_{n-1,n+1}+b^{(1)}_{n,n+1}+b^{(2)}_{n-1,n}+b^{(2)}_{n-1,n+1}\Big]=Ea_{n-1,n,n+1},
\end{equation}
and
\begin{eqnarray}
&&(4-J)b^{(1)}_{n,n+1}-b^{(1)}_{n,n+2}-Jb^{(1)}_{n-1,n+1}-b^{(2)}_{n,n+1}+(J-1)a_{n-1,n,n+1}=Eb^{(1)}_{n,n+1},\nonumber\\
&&(4-J)b^{(2)}_{n-1,n}-b^{(2)}_{n-2,n}-Jb^{(2)}_{n-1,n+1}-b^{(1)}_{n-1,n}+(J-1)a_{n-1,n,n+1}=Eb^{(2)}_{n-1,n}.
\end{eqnarray}

The following substitution
\begin{eqnarray}
&&a_{n_1,n_2,n_3}(k_1,k_2,k_3)=\sum_{a,b,c=1}^3\varepsilon_{abc}A(k_a,k_b)A(k_a,k_c)A(k_b,k_c){\rm e}^{i(k_an_1+k_bn_2+k_cn_3)},
\nonumber\\
&&b^{(1)}_{n_1,n_2}(k_1,k_2,k_3)=\frac{1}{2}\sum_{a,b,c=1}^3\varepsilon_{abc}B(k_a,k_b)A(k_a,k_c)A(k_b,k_c){\rm e}^{i[(k_a+k_b)n_1+k_cn_2]},\nonumber\\
&&b^{(2)}_{n_1,n_2}(k_1,k_2,k_3)=\frac{1}{2}\sum_{a,b,c=1}^3\varepsilon_{abc}A(k_a,k_b)A(k_a,k_c)B(k_b,k_c){\rm e}^{i[k_an_1+(k_b+k_c)n_2]}.
\end{eqnarray}
where $A(k,\tilde k)$ and $B(k,\tilde k)$ are given by equations (29) and (30),
solves equations (32)-(34) giving the energy
\begin{equation}
E(k_1,k_2,k_3)=\sum_{j=1}^3E_{magn}(k_j).
\end{equation}

At the same time equation (35) turns into
\begin{equation}
X^{(j)}(k_1,k_2,k_3){\rm e}^{i(k_1+k_2+k_3)n}=0,\qquad j=1,2,
\end{equation}
where
\begin{eqnarray}
&&X^{(1)}(k_1,k_2,k_3)=\frac{1}{2}\sum_{a,b,c=1}^3\varepsilon_{abc}{\rm e}^{ik_c}A(k_a,k_c)\Big[\Big(E(k_1,k_2,k_3)-4+{\rm e}^{ik_c}
+J\Big(1+{\rm e}^{-i(k_a+k_b)}\Big)\Big)\nonumber\\
&&\cdot B(k_a,k_b)A(k_b,k_c)
+{\rm e}^{ik_b}A(k_a,k_b)B(k_b,k_c)+2(1-J){\rm e}^{-ik_a}A(k_a,k_b)A(k_b,k_c)\Big],\nonumber\\
&&X^{(2)}(k_1,k_2,k_3)=\frac{1}{2}\sum_{a,b,c=1}^3\varepsilon_{abc}{\rm e}^{-ik_a}A(k_a,k_c)\Big[\Big(E(k_1,k_2,k_3)-4+{\rm e}^{-ik_a}
+J\Big(1+{\rm e}^{i(k_b+k_c)}\Big)\Big)\nonumber\\
&&\cdot A(k_a,k_b)B(k_b,k_c)
+{\rm e}^{-ik_b}B(k_a,k_b)A(k_b,k_c)+2(1-J){\rm e}^{ik_c}A(k_a,k_b)A(k_b,k_c)\Big].
\end{eqnarray}

An evaluation of the sums in (39) with the use of the computer algebra system MAPLE gives
\begin{eqnarray}
&&X^{(1)}(k_1,k_2,k_3)=(1-J^2)\varphi(k_1,k_2,k_3)\Big[\Big(E(k_1,k_2,k_3)-5\Big){\rm e}^{ik}-1+J\Big(2+3{\rm e}^{ik}+{\rm e}^{2ik}\Big)\Big],\nonumber\\
&&X^{(2)}(k_1,k_2,k_3)=(1-J^2)\varphi(k_1,k_2,k_3)\Big[{\rm e}^{ik}-E(k_1,k_2,k_3)+5-J\Big(3+2{\rm e}^{ik}+{\rm e}^{-ik}\Big)\Big].
\end{eqnarray}
where $k=k_1+k_2+k_3$ and
\begin{equation}
\varphi(k_1,k_2,k_3)=\Big({\rm e}^{ik_1}-{\rm e}^{ik_2}\Big)\Big({\rm e}^{ik_2}-{\rm e}^{ik_3}\Big)\Big({\rm e}^{ik_3}-{\rm e}^{ik_1}\Big)
\cdot\prod_{j=1}^3(1-{\rm e}^{ik_j}).
\end{equation}

As we see from (40) $X^{(j)}(k_1,k_2,k_3)\equiv0$, $(j=1,2)$ only in two integrable cases \cite{13,15} $J=1$ and $J=-1$.
However even in the general case the whole system (32)-(35) will be obviously satisfied for the following wave functions ($j=1,2$)
\begin{eqnarray}
a_{n_1,n_2,n_3}(k_1,k_2,k_3,\tilde k_1,\tilde k_2,\tilde k_3)=\varphi(\tilde k_1,\tilde k_2,\tilde k_3)a_{n_1,n_2,n_3}(k_1,k_2,k_3)-
\varphi(k_1,k_2,k_3)a_{n_1,n_2,n_3}(\tilde k_1,\tilde k_2,\tilde k_3),\nonumber\\
b^{(j)}_{n_1,n_2}(k_1,k_2,k_3,\tilde k_1,\tilde k_2,\tilde k_3)=\varphi(\tilde k_1,\tilde k_2,\tilde k_3)b^{(j)}_{n_1,n_2}(k_1,k_2,k_3)-
\varphi(k_1,k_2,k_3)b^{(j)}_{n_1,n_2}(\tilde k_1,\tilde k_2,\tilde k_3),
\end{eqnarray}
where
\begin{equation}
\prod_{l=1}^3{\rm e}^{i\tilde k_l}=\prod_{l=1}^3{\rm e}^{ik_l},\quad\sum_{l=1}^3E_{magn}(\tilde k_l)=\sum_{l=1}^3E_{magn}(k_l).
\end{equation}

\section{Remark on the four-magnon problem}

An evaluation of four-magnon states for our model is a problem of the next level of complexity. In order to see this let us recall an evaluation of equation (42). First of all we take a triple $\{k_1,k_2,k_3\}$ and then construct the wave function (36) which satisfies equations (32)-(34) but does not satisfies equation (35). In order to solve the latter we add the  term related to a new triple $\{\tilde k_1,\tilde k_2,\tilde k_3\}$. The resulting wave function has the form (42).

For a four-magnon state
\begin{eqnarray}
&&|4\rangle=\Big[\sum_{n_1<n_2<n_3<n_4}a_{n_1,n_2,n_3,n_4}{\bf S}_{n_1}^-{\bf S}_{n_2}^-{\bf S}_{n_3}^-{\bf S}_{n_4}^-
+\sum_{n_1<n_2<n_3}\Big(b^{(1)}_{n_1,n_2,n_3}({\bf S}_{n_1}^-)^2{\bf S}_{n_2}^-{\bf S}_{n_3}^-\nonumber\\
&&+b^{(2)}_{n_1,n_2,n_3}{\bf S}_{n_1}^-({\bf S}_{n_2}^-)^2{\bf S}_{n_3}^-+b^{(3)}_{n_1,n_2,n_3}{\bf S}_{n_1}^-{\bf S}_{n_2}^-({\bf S}_{n_3}^-)^2\Big)+\sum_{n_1<n_2}c_{n_1,n_2}({\bf S}_{n_1}^-)^2({\bf S}_{n_2}^-)^2\Big]|\emptyset\rangle.
\end{eqnarray}
the corresponding ${\rm Schr\ddot odinger}$ equation should again split on several systems of equations related to different processes in the four-magnon system. We shall study only one of them related to an extreme right magnon being separated from the others and hence do not interacting with them.

As in the previous case we take at once $a_{n_1,n_2,n_3,n_4}$ as a linear combination of $24=4!$ Bethe exponents related to different permutations of four wave numbers $k_1$, $k_2$, $k_3$ and $k_4$. Let us first consider six of them proportional to ${\rm e}^{ik_4n_4}$. From an account of the interaction between the triple of left magnons, it follows
that these six terms should be added to another six ones proportional to the same exponent ${\rm e}^{ik_4n_4}$ but with a new triple $\{\tilde k_1,\tilde k_2,\tilde k_3\}$ of the left magnons wave numbers. The same picture will be for all other three groups of exponents proportional to ${\rm e}^{ik_jn_4}$ ($j=1,2,3$).

Now we may explain the cardinal difference between three- and four-magnon problems. In the former case for a given triple $\{k_1,k_2,k_3\}$ it is sufficient to add only a single triple $\{\tilde k_1,\tilde k_2,\tilde k_3\}$, however in the latter one for a given quartet $\{k_1^{(0)},k_2^{(0)},k_3^{(0)},k_4^{(0)}\}$ it is necessary to add at least four different new quartets
\begin{equation}
\{k_1^{(j)},\dots,k_4^{(j)}\},\qquad k^{(j)}_j=k_j^{(0)},\qquad j=1,\dots,4,
\end{equation}
related the same energy and total wave number. Each of the four induced quartets has the similar rights to the initial one and hence should be in the same correspondence with some other four quartets (one of them is the initial quartet). As a result the total set of quartets may be represented as a graph whose {\it each} vertex (related to its own quartet) is connected with four different other vertices.

Let us study in detail a construction of the simplest example of such set of quartets.
As it was explained above, first of all we take an initial quartet $\{k_1^{(0)},\dots,k_4^{(0)}\}$ which induces the four new ones according to equation (45).
Taking now the quartet $\{k_1^{(1)},\dots,k_4^{(1)}\}$ and applying the same argumentation as for $\{k_1^{(0)},\dots,k_4^{(0)}\}$ we see that it also must be connected with four different quartets. One of them is already known: it is $\{k_1^{(0)},\dots,k_4^{(0)}\}$. Hence, we should present the rest three quartets. The simplest way to do this is to use the quartets (45) with $j=2,3,4$ (otherwise we have to introduce new quartets $\{k_1^{(j)},\dots,k_4^{(j)}\}$, $j=5,6,7$). Under this choice any pair $\{k_1^{(1)},\dots,k_4^{(1)}\}$ and $\{k_1^{(j)},\dots,k_4^{(j)}\}$ ($j=2,3,4$) should have a common wave number. According to equation (45) the wave numbers $k_j^{(j)}$ ($j=1,\dots,4$) are already utilized. Hence without lost of generality we may put $k_j^{(1)}=k_1^{(j)}$ ($j=2,3,4$). Turning to the quartet $\{k_1^{(2)},\dots,k_4^{(2)}\}$ we see that since it has been already connected with $\{k_1^{(0)},\dots,k_4^{(0)}\}$ and $\{k_1^{(1)},\dots,k_4^{(1)}\}$ we have to connect it only with two quartets. Again, the simplest way to do this is to use $\{k_1^{(j)},\dots,k_4^{(j)}\}$ ($j=3,4$). Since the wave numbers $k_j^{(2)}$ with $j=1,2$ are already utilized we may postulate (without any lost of generality) $k_j^{(2)}=k_2^{(j)}$ ($j=3,4$). Finitely we consider the quartet $\{k_1^{(3)},\dots,k_4^{(3)}\}$ and connect it with $\{k_1^{(4)},\dots,k_4^{(4)}\}$ by the formula $k_4^{(3)}=k_3^{(4)}$. As a result we have the system of five quartets
(geometrically it may be represented as a graph with five vertices connected to each other)
\begin{equation}
\{k_1^{(j)},\dots,k_4^{(j)}\},\qquad j=0,\dots,4,
\end{equation}
and 10 relations; namely (45) and
\begin{equation}
k^{(j)}_l=k^{(l)}_j,\qquad j,l=1,\dots,4,\quad j\neq l.
\end{equation}
According to Eqs. (45) and (47) only 10 of the 20 wave numbers $k^{(j)}_l$ ($j=0,\dots4,\quad l=1,\dots4$) for example $k^{(j)}_l$ ($0<j\leq l,\quad l=1,\dots4$) are independent. According to the energy and quasimomentum conservation laws they satisfy 10 independent equations
\begin{eqnarray}
&&{\rm e}^{i(k^{(1)}_1+k^{(2)}_2+k^{(3)}_3+k^{(4)}_4)}={\rm e}^{i(k^{(1)}_1+k^{(1)}_2+k^{(1)}_3+k^{(1)}_4)}=
{\rm e}^{i(k^{(1)}_2+k^{(2)}_2+k^{(2)}_3+k^{(2)}_4)}\nonumber\\
&&={\rm e}^{i(k^{(1)}_3+k^{(2)}_3+k^{(3)}_3+k^{(3)}_4)}={\rm e}^{i(k^{(1)}_4+k^{(2)}_4+k^{(3)}_4+k^{(4)}_4)}={\rm e}^{ik},
\end{eqnarray}
and
\begin{eqnarray}
&&\sum_{j=1}^4E_{magn}(k^{(j)}_j)=\sum_{j=1}^4E_{magn}(k^{(1)}_j)=E_{magn}(k^{(1)}_2)+\sum_{j=2}^4E_{magn}(k^{(2)}_j)
\nonumber\\
&&=\sum_{j=1}^3E_{magn}(k^{(j)}_3)+E_{magn}(k^{(3)}_4)=\sum_{j=1}^4E_{magn}(k^{(j)}_4)=E.
\end{eqnarray}

Correspondingly a general system of $M$ quartets contains $4M$ wave numbers. Since each of them is common to two different quartets, only a half of them (namely $2M$) are independent. Equations
\begin{equation}
\prod_{l=1}^4{\rm e}^{ik_l^{(j)}}={\rm e}^{ik},\quad\sum_{l=1}^4E_{magn}(k_l^{(j)})=E,\qquad j=1,\dots M.
\end{equation}
where $k$ is the quasimomentum (total wave number) and $E$ is the energy, give $2M$ conditions on these $2M$ wave numbers.
Hence the existence of finite-$M$ four-magnon Bethe wave functions is in question even without an analysis of the pure four-magnon collisions.

\section{Summary and discussion}

In the present paper we have studied the three-magnon problem for a general isotropic $S=1$ ferromagnetic infinite chain. Except the two integrable
cases \cite{13,15} the corresponding wave functions can not be represented in the Bethe form (7) but only as a non-integrable modification (9), (42) (we call it the degenerative, discrete-diffractive form).
Since the presented set of states is highly overloaded a complete description of the three-magnon scattering \cite{16} may be obtained only after an extraction of a non-overloaded complete system of the three-magnon states. However, it is not clear how to represent such a system. In fact, for an integrable spin chain the three-magnon eigenstates may be parameterized by their energy $E$, quasimomentum $k$ and the eigenvalue of an additional first integral. The latter belongs to an infinite set of commuting first integrals which may be obtained by the standard procedure \cite{1,2,3,15}. The system studied in the present paper is however non-integrable. Probably there exist an operator
$\hat Q$ which commutes both with the Hamiltonian and the shift operator. If its eigenvalue ${\cal Q}$ is independent from $E$ and $k$, it may be used for a parametrization of the spectrum. Otherwise the parametrization procedure seems unclear and probably may be developed on the base of noncommutative geometry \cite{17}.
Nevertheless, it seems evident that a scattering of three incoming magnons with wave numbers $\{k_1,k_2,k_3\}$ should
result in the creation of all possible outgoing three-magnon states with the same $E$ and $k$.

We also have shown that the corresponding four-magnon problem is much more difficult.

We suggest that the obtained result in its future development may be useful for a derivation of low-temperature expansions for thermodynamical quantities in the gapped regime \cite{18,19,20}.

Finitely we notice that although equation (4) on the $S$-matrix has the same form as the equation on the so called $R$-matrix (the Yang-Baxter equation in the braid group form \cite{1,2,7,15}) the two subjects are not directly connected to each other. The $S$-matrix characterizes a two-magnon scattering and its dimension $d^2\times d^2$ depends on the number of elementary excitations (that is $d$ in our notations). From the other hand the $R$-matrix has the same dimension
$\tilde d^2\times\tilde d^2$ as the Hamiltonian density matrix. Here $\tilde d$ is the dimension of the Hilbert space associated with each site of the chain (for example $\tilde d=4$ for spin ladders \cite{5} and $\tilde d=3$ for the $t-J$ model \cite{6}). Usually $d<\tilde d$. Moreover as it was shown in the present paper the
Yang-Baxter equation for the $S$-matrix may be satisfied even in non-integrable cases when the $R$-matrix
formalism is irrelevant.

The author is grateful to L. D. Faddeev for the helpful discussion and to H. Katsura for the useful comment.

\end{document}